# A Guide of Fingerprint Based Radio Emitter Localization using Multiple Sensors

Tao YU[†], Azril HANIZ[†], *members,* Kentaro SANO[†], Ryosuke IWATA[†],
Ryouta KOSAKA[†], Yusuke KUKI[††], *nonmembers,* Gia Khanh TRAN[†], *member,*
Jun-ichi TAKADA[†], *fellow, and* Kei SAKAGUCHI[†], *senior member*

**SUMMARY** Location information is essential to varieties of applications. It is one of the most important context to be detected by wireless distributed sensors, which is a key technology in Internet-of-Things. Fingerprint-based methods, which compare location unique fingerprints collected beforehand with the fingerprint measured from the target, have attracted much attention recently in both of academia and industry. They have been successfully used for many location-based applications. From the viewpoint of practical applications, in this paper, four different typical approaches of fingerprint-based radio emitter localization system are introduced with four different representative applications: localization of LTE smart phone used for anti-cheating in exams, indoor localization of Wi-Fi terminals, localized light control in BEMS using location information of occupants, and illegal radio localization in outdoor environments. Based on the different practical application scenarios, different solutions, which are designed to enhance the localization performance, are discussed in detail. To the best of the authors' knowledge, this is the first paper to give a guideline for readers about fingerprint-based localization system in terms of fingerprint selection, hardware architecture design and algorithm enhancement.
*key words: Internet-of-Things, wireless distributed sensors, radio emitter localization, fingerprint, different applications, guideline, prototyping*

## 1. Introduction

Wireless Distributed Sensors (WDS) play important roles for monitoring status/context of environments/users in diverse applications of Internet-of-Things (IoT) [1-4]. One of the important contexts to be detected by WDS is location information of users/devices to be used in many applications such as BEMS (Building Energy Management System), logistics, marketing, mobility, and smart society, in addition to agriculture [5-11].

Location of users/devices can be detected by measuring radio waves emitted from users/devices with multiple sensors. For instance, location of users can be measured directly by detecting infrared signals emitted by people with human detection sensors [12][13], or by detecting light signals reflected by users with image sensors [14-16], or it can be estimated indirectly by measuring radio waves emitted by users such as from smart phones [17][18]. Sensing device should be designed carefully based on the target applications, scenarios, and cost. For example, infrared human detection sensors might be selected in applications of BEMS due to the efficiency of cost, while expensive wideband radio receivers might be designed in applications of illegal radio surveillance due to the supporting coverage of space and frequency to be measured.

Traditionally, geometry based methods, such as multilateration by measuring TOF (Time-of-Flight) between the target and multiple sensors or triangulation by measuring AOA (Angle-of-Arrival) of the target signals at multiple sensors [19], have been used for localization. However, since these methods rely on the LOS (Line-of-Sight) signal from the target, it is difficult to apply them to the environment with many multipath signals such as in indoors or urban city environments. On the other hand, fingerprint-based methods, which compare location unique fingerprints collected beforehand with the fingerprint of the target, have attracted much attention recently due to the innovation of statistics, especially machine learning [20]. Here, fingerprints are defined as location-specific data sets of the propagation channel such as RSSI (Received Signal Strength Indicator) including the effect of multi-paths, and the target location is estimated via pattern matching algorithm using the fingerprint database. The offline collection of fingerprints can be considered as a supervised learning process to develop the fingerprint database with indices of known locations. Since it is impossible to measure the fingerprints in all the space and frequency, statistical regression algorithms interpolate the data sets by using the property of propagation channel in space and frequency [21]. In the phase of pattern matching, statistical tracking algorithms such as the Kalman filter, Particle filter, and Recurrent Neural Network (RNN) will enhance the performance by exploiting prior knowledge of users [22].

A majority of literatures on fingerprint-based methods focus on indoor localization applications. The RADAR system [23], which is one of the pioneering works in fingerprint-based localization, proposed to utilize RSSI fingerprints obtained from indoor WLAN systems to perform localization and tracking. In [24], RSSI fingerprints obtained from a WiMAX network was utilized to perform outdoor localization in an urban area. As RSSI fingerprints only indicate the instantaneous power

**Table 1** Four different approaches of fingerprint based radio emitter localization for four different applications.

| Section | Scenario | Application | Target Signal | Target Location | Selected fingerprint | Technology | Requirement on accuracy | Cost | Complexity² |
|---|---|---|---|---|---|---|---|---|---|
| 3 | Indoor | Cheating detection | Known | Discrete | CIR | Baseline | ≈ Distance between seats | Expensive | $O(LN_{ant}^2 N_{gr})$ |
| 4 | Indoor | Indoor navigation | Known | Continuous | RSSI + RSPD | Tracking using particle filters | ≈ Distance between room objects (e.g. one meter) | Cheap | $O(N_{ant}^2 MP)$ |
| 5 | Indoor | Light control | Known | Continuous | Human detection | Tracking using Kalman filter, RNN (Recurrent Neural Network) | ≈ Distance between office lights (e.g. few meters) | Very cheap | $O\left((M+N_{gr})N_{gr}\right)$ |
| 6 | Outdoor | Illegal radio | Unknown | Continuous | Cross-correlation of Rx signals & phase-difference | Interpolation & regression | ≈ Distance between vehicles (e.g. few tens of meters)¹ | Very expensive | $O(M(L+N_{ant}^2)N_{gr})$ + cost of interpolation³ |

1 Based on the statistics of illegal radios in Japan [51], a significant portion consists of illegal radios used by truck drivers. Thus, the localization system should be able to distinguish between large vehicles.
2 Only the online phase complexity is considered here because computations in the offline phase does not affect real-time performance of the algorithms.
3 Computational cost of interpolation in multiple domains is high, and more details can be found in [48] [49].

levels of received signal but fail to capture channel's temporal characteristics such as the delay spread caused by multipaths, many researchers have proposed algorithms that utilize CSI (Channel State Information) fingerprints, which contain rich channel frequency response information and can be extracted from WLAN systems. For example, the PinLoc system was proposed for indoor localization in crowded university buildings as well as in a museum [25]. In [26], an indoor localization system utilizing CSI obtained from a multiple antenna WLAN system was proposed, and it could successfully incorporate spatial information into the CSI fingerprint to improve accuracy. The CIR (Channel Impulse Response) fingerprint was also utilized in [27] to locate and track mobile terminals in an underground mine, and the fingerprints were measured using a vector network analyzer. A comprehensive survey regarding fingerprint-based algorithms can be found in [28][29].

In this paper, four different approaches of fingerprint based radio emitter localization are introduced for four different applications as summarized in Table 1. Types of fingerprint, hardware architecture of sensors, and pattern matching algorithm are selected based on the conditions of applications. Application-specific requirements on accuracy and cost are also summarized and categorized in the table. The first application is localization of LTE smart phone used for anti-cheating detection in exams [30], the second application is indoor localization of Wi-Fi terminals [31], the third application is light control in BEMS using location information of people [32], and the last application is illegal radio localization in outdoor environments [33]. Based on the different given conditions in different applications such as knowledge of target signals, different pattern matching algorithms can be designed to enhance the localization performance.

To the best of the authors' knowledge, this work is the first paper to give a guideline for readers about fingerprint-based localization in terms of the following three points.

1. How to select type of fingerprint based on the given conditions of applications.
2. How to design hardware architecture of sensors to measure the selected fingerprint.
3. How to enhance pattern matching algorithm based on the given conditions of applications.

The readers might follow the flowchart in Fig. 1 to decide a proper fingerprint-based localization method to realize their applications of interest. First, users must decide whether their approach is fingerprint-based or not. Since the fingerprint-based approach in this paper is based on database measured in advance, if users want an on-the-fly localization method without database, please select conventional methods, e.g. triangulation. Otherwise, please follow our methods in the next step. Next, please confirm whether the type of the system is an infrastructure-based method or self-localization method, e.g. SLAM (Simultaneous localization and mapping). Self-localization schemes based on sensors' self-measured fingerprint are out of the scope of this paper. In this paper, fingerprints measured from multi-sensors are gathered at an infrastructure, called fusion center, to facilitate the pattern matching step in the estimation phase. After that, please confirm the type of sensing signals that your sensors provide. Sonar signals are out of the scope of this paper. However, since sonar signals follow similar physical properties as those of radio waves, the approach in Sect. 3 might be helpful. If the sensor's output is a visual image, please refer to the approach in Sect. 5. Otherwise, Sect. 3, 4 and 6 are suitable for radio-wave fingerprints. In the next step, please confirm your knowledge about the target emitter. If information about the target's signal, e.g. center frequency and bandwidth of illegal emitters, is unknown, please go to Sect. 6. Otherwise, Sect. 3 and 4 are suitable for localization of legal emitters. The difference between these two sections is about the investment cost and is inversely proportional with required accuracy for the infrastructure. Section 3 is suitable for application-specified custom made systems while Sect. 4 employs low-cost commercialized hardware. The authors hope that this guideline can cover most of types of fingerprint-based localization methods for any purpose of the readers.

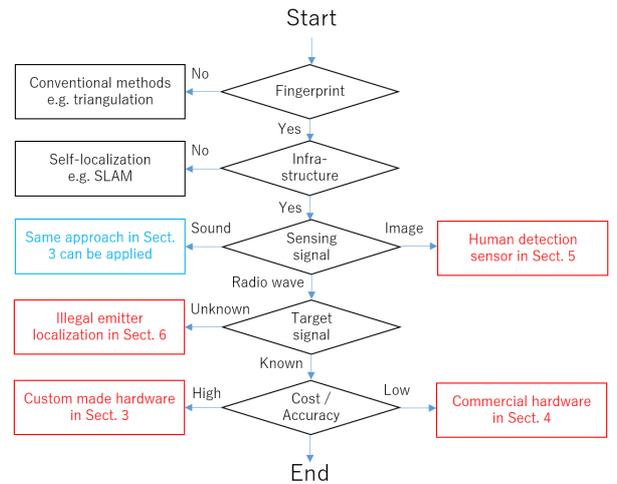

**Fig. 1** A guideline for fingerprint-based localization methods.

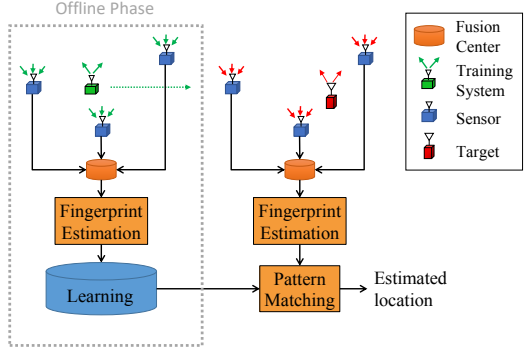

**Fig. 2** System architecture for fingerprint-based localization

The rest of this paper is organized as follows. Section 2 provides the common mathematical background of fingerprint-based localization to be used in all applications. Section 3 describes localization of smart phones emitting LTE sounding signals as the first application. Sect. 4 describes the second application of localization of Wi-Fi devices using Wi-Fi training preambles, Sect. 5 provides the third application of localization of moving people detected by battery-less human detection sensors, and Sect. 6 introduces the last application of localization of radio terminals emitting illegal radio signals. Finally, Sect. 7 concludes this paper.

## 2. Background on Fingerprint-based Localization

In this section, we shall provide some fundamental background knowledge regarding fingerprint-based localization methods in general, which will form the basic framework for the following sections. Fingerprints can be defined as location-specific data sets or unique spatial signatures, and by tagging these fingerprints with geographical coordinates, the locations are able to be identified and differentiated.

Without loss of generality, in this paper the authors shall denote the fingerprint vector as $\mathbf{F} = [F_1, F_2, \dots, F_N]^T$, where $N$ denotes the dimensionality of the fingerprints, which might be the number of snapshots or sensors depending on the algorithm. By extending the dimensionality of this vector, it is expected that the uniqueness of the fingerprints over space will also increase, resulting in a better localization performance. This can be achieved by increasing the number of sensors. However, combining several types of fingerprints may be more advantageous as we can exploit the strengths of each type of fingerprint simultaneously.

A common system architecture for fingerprint-based algorithms is shown in Fig. 2. Several sensors are placed covering the area of interest, and they are connected to the fusion center by a backhaul network, which may be either using a wired connection or wirelessly. The fusion center operates by collecting data measured at every sensor, and it may also send commands to the sensors. Generally, fingerprint-based methods comprise of two phases. The first phase is called the offline phase, or also known as the learning or training phase, and the second phase is called the online or estimation phase.

In the offline phase, fingerprints from many locations are estimated using data from several sensors. In practice, due to numerous constraints, fingerprints can only be collected discretely over space from a limited number of locations, and this results in a finite number of fingerprints being stored in a database together with their geographical coordinates. In the online phase, localization of the target is performed by first estimating the target's fingerprint, and then performing pattern matching with the fingerprint stored in the database. There are several approaches for performing pattern matching, and among them a simple way would be to minimize the Euclidean distance, which can be expressed as [23]:

$$\arg\min_i \|\mathbf{F}^{\text{target}} - \mathbf{F}_i^{\text{DB}}\| \qquad (1)$$

where $\mathbf{F}^{\text{target}}$ denotes the target's fingerprint vector, and $\mathbf{F}_i^{\text{DB}}$ denotes the fingerprint in the database associated with coordinates $\mathbf{u}_i$. We may also employ a probabilistic approach, which aims to maximize the posterior distribution. Utilizing the Bayes' theorem, it can be expressed as

$$p(\mathbf{u}|\mathbf{F}^{\text{target}}) = p(\mathbf{F}^{\text{target}}|\mathbf{u})p(\mathbf{u})/p(\mathbf{F}^{\text{target}}) \qquad (2)$$

where the denominator is usually replaced with a constant. If the prior distribution of the target's location $p(\mathbf{u})$ is unknown, we may use approximation or assume a uniform distribution instead, and maximizing the posterior distribution will be similar to maximizing likelihood $L(\mathbf{u}) = p(\mathbf{F}^{\text{target}}|\mathbf{u})$, which is the basis for maximum likelihood estimation (MLE). However, since the fingerprints are discrete over space, we should utilize the following approximation.

$$\arg\max_{\mathbf{u}} L(\mathbf{u}) \approx \arg\max_{\mathbf{u}_i} p(\mathbf{F}^{\text{target}}|\mathbf{u}_i) \qquad (3)$$

The likelihood function at each training location $p(x|\mathbf{u}_i)$ can be modelled by some distribution such as the Gaussian distribution, and the distribution parameters are learned by utilizing the collected fingerprints during the offline phase [31]. However, we can further improve localization performance by introducing prior knowledge such as previous location estimates into $p(\mathbf{u})$, and this is the basis for dynamic tracking algorithms such as the Kalman filter and particle filters [24].

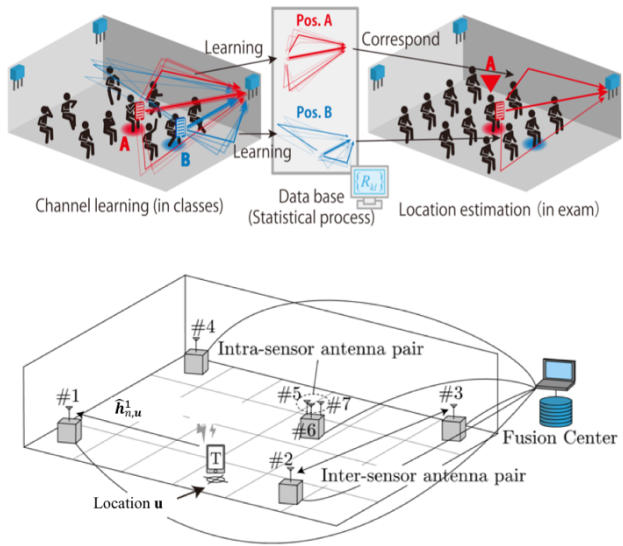

**Fig. 3** Illustration of system structure

In the offline phase, we have discussed so far about pattern matching based on fingerprints collected over discrete space, but in practice, the target may be located in non-discrete space which is not restricted to the locations of the collected fingerprints. It would be necessary to infer or predict the fingerprint at these "holes" in the database. If the underlying model of the fingerprint is known, expanding the discrete fingerprint database would be straightforward. However, in most cases, the assumed model is imperfect, or no prior knowledge of the fingerprint model is available, thus the model would have to be learned from the measured data.

One approach is employing regression techniques to find the relationship between location and fingerprint, or the trend of fingerprints over discrete space. In general, this can be expressed as $\mathbf{F} \approx f(\mathbf{u}, \boldsymbol{\beta})$, where $f(\cdot)$ denotes the regression function, and $\boldsymbol{\beta}$ denotes a set of parameters which are learned from the data. An example of applying regression to discretely sampled spatial data is the Kriging algorithm [34], which is also known as Gaussian Process Regression. Here, the data across the spatial domain is assumed to be generated by a Gaussian process, and regression is performed by learning the spatial covariance from measured data. Another example is regression along the frequency domain, where a log-linear model with respect to frequency is employed to predict fingerprints at any frequency [35]. In these given examples, the Gaussian distribution function and the log-linear function can be regarded as kernel functions. This learning concept is not limited to fingerprint models, but it can also be utilized for learning the model of sensor measurements, such as the coverage of motion sensors. This learning step in the offline phase can also be conducted using machine learning algorithms, such as the Support Vector Machine (SVM) and neural networks, which may be more effective identifying subtle patterns hidden in the measured data or predicting target movement in tracking applications.

In conventional non-fingerprint-based algorithms, the relationship between location and fingerprint can be expressed in a closed form equation, which enables us to calculate the theoretical bounds for localization accuracy, such as the Cramer Rao Lower Bound (CRLB) [36]. However, it should be noted that this is not the case for fingerprint-based algorithms. The accuracy not only depends on the system parameters such as the density of fingerprint database and type of fingerprint, but also largely depends on the surrounding environment. Therefore, the localization accuracy will greatly differ depending on the application of the system.

## 3. Localization of smart phones emitting LTE sounding signals

In this section, a localization system of smart phones emitting LTE sounding signals is described. In [30], the authors proposed a method for location estimation of a cell-phone in indoor environments for anti-cheating detection in examinations. The proposed method is a location fingerprint scheme which employs the statistical characteristics of the signal cross-correlation among multiple sensors. With the prior knowledge of SRS (Sounding Reference Signal), the cross-correlations between different pairs of CIR in antennas are employed as fingerprint [37]. The proposed method can be considered as a generalized scheme of conventional location fingerprint schemes. Besides, compared with conventional methods, in which fine grid location measurements are required and environment must be static, the proposed one invokes statistical learning technique and estimates location based on the correlation of received samples with the statistical learning database, so the proposed method has superior estimation accuracy and installation simplicity.[37]

### 3.1 System structure

The illustration of the localization system is shown in Fig. 3. Sensors equipped with one antenna are placed in the target environment, e.g., a class room. There was another sensor equipped with a 3-antenna array located at the center of the class room. All sensors are directly connected to a fusion center via RF cables, and the fusion center is monitoring the SRSs transmitted in the environment all the time. The transmitter to be detected is an LTE cell-phone, which sometimes is used for cheating in examinations. In the learning phase, the transmitter was respectively put at each measuring point corresponding to a seat of the class room. The grid spacing is sufficiently larger than the wave-length which means there is no spatial correlation between each grid point. In the proposed method, by fitting the fingerprint distribution into Gaussian distribution, high accuracy can be achieved, despite of rough grid of measurements with large spatial intervals. This differs the proposed method from the conventional location fingerprint methods in which fingerprints are collected on fairly dense grid measurements in order to improve accuracies. The detailed algorithm and parameters will be given respectively in Sec. 3.2 and 3.3.

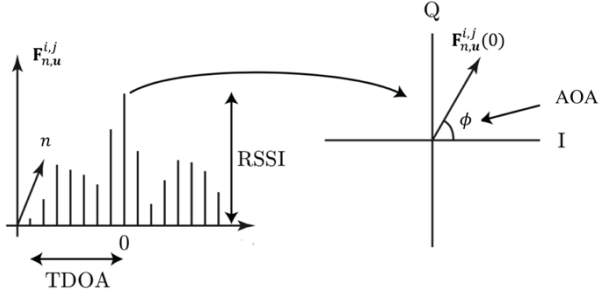

**Fig. 4** An example of cross-correlation of CIR

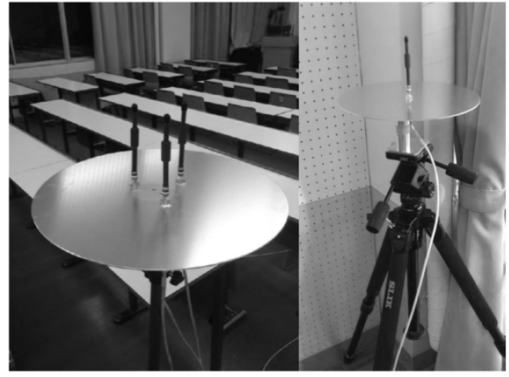

**Fig. 5** Sensor equipped with 3-antenna array located at center (left) and single antenna sensor located at four corners (right) of the class room.

### 3.2 Proposed localization algorithm

In LTE, SRS is transmitted by the UE using a known sequence and used to estimate the uplink channel quality. The SRS has constant amplitude zero autocorrelation (CAZAC) property, since it is derived from the Zadoff-Chu sequences. Therefore, the authors calculate the sliding cross-correlation between a replica of pre-known SRSs and received signal to search the target signal, and extract the CIR information. The proposed algorithm has two phases: learning phase and estimation phase.

*A. Learning phase.*

Define $\mathbb{C}[a, b]$ as the cross-correlation between vector $a$ and $b$. The employed fingerprint is defined as the cross-correlation between the CIRs at different antenna pairs:

$$F_{n,u}^{i,j} = \mathbb{C}[\hat{h}_{n,u}^i, \hat{h}_{n,u}^j] \qquad (4)$$

where $\hat{h}_{n,u}^j \in C^L$ is the $n$-th snapshot of estimated CIR in antenna $j$ when user is located in $u$. $L$ is the maximum delay tap of the CIR to be considered. Obviously $F_{n,u}^{i,j} \in C^{2L}$. Fig. 4 gives an example of the cross-correlation of a pair of CIRs. It shows that the proposed fingerprint can generalize conventional location fingerprint methods in terms of using RSSI, TDOA and AOA at the same time.

To learn the cross-correlation vector $F_{n,u}^{i,j}$, in this paper, the authors use a MLE with the assumption that the probabilistic distribution of $F_{n,u}^{i,j}$ follows a multi-dimensional Gaussian distribution. Thus the fingerprint's probability function forming decision region is defined as:

$$p(F^{i,j}|u) = \frac{1}{(2\pi)^{2L} \det\left(\hat{R}_u^{i,j}\right)} \qquad (5)$$

$$\exp\left[-\frac{1}{2}(F^{i,j} - \hat{F}_u^{i,j})^H \left(\hat{R}_u^{i,j}\right)^{-1} (F^{i,j} - \hat{F}_u^{i,j})\right]$$

where the empirical parameters $\hat{F}_u^{i,j}$ and $\hat{R}_u^{i,j}$ are the estimated mean vector and covariance matrix of $F_{n,u}^{i,j}$ respectively.

*B. Estimation phase*

In the estimation phase, when the sensors detect suspicious signals sent from an unknown position in the exam, the proposed method starts to estimate the location of the cell-phone for anti-cheating detection. In the learning phase, the system employing the proposed method has abstained the learning database. Therefore, the most suspicious location can be estimated as the location with the highest likelihood which is computed by substituting cross-correlation of the extracted CIR $F^{i,j}$ from the target into Eq. (5) for all location candidates $u$. Now the authors assume the prior probability about $u$ is uniform and the reliability of each antenna pair is equal. Then, the log-likelihood function can be calculated as in Eq. (3) and the estimated location $\hat{u}$ can be derived:

$$\hat{u} = \underset{u}{\mathrm{argmax}} \sum_i \sum_j \log\{p(F^{i,j}|u)\} \qquad (6)$$

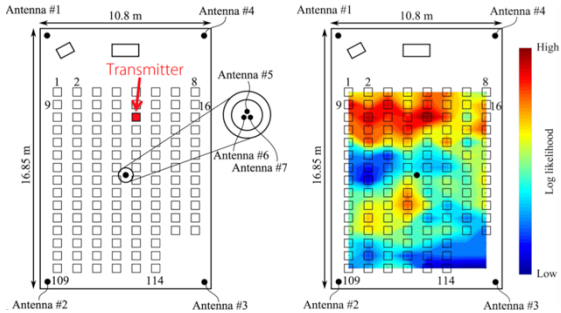

**Fig. 6** Environment of experiment and example of estimated likelihood map.

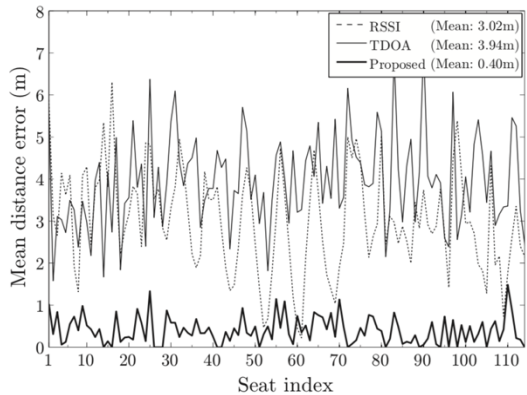

**Fig. 7** Mean distance error of each measurement point

Maximization of the likelihood above is performed using a brute-force search over $N_{gr}$ grid points, and this results in complexity $O(LN_{ant}^2 N_{gr})$, where $N_{ant}$ denotes the number of antennas. It is derived based on the fact that each fingerprint has $2L$ dimensions, and is calculated for every antenna pair.

### 3.3 Experiment and results

An experiment was conducted to validate the proposed method in a class room at Tokyo Institute of Technology simulating an entrance exam with a cheating scenario using a cell-phone. Four sensors each equipped with one antenna were located at every corner of the class room, and another sensor equipped with a 3-antenna array was located at the center of the class room, as is shown in Fig. 5. The transmitter was an LTE cell-phone with the center frequency $f_c = 1.9575$ GHz and a bandwidth B = 3.6 MHz. The transmitter was respectively set at each grid point corresponding to a seat.

Nobody was in the room except one experiment staff, so the multi-path environment was nearly constant. At seat, the transmitter was respectively placed every 2cm and measurements were performed twice at each place, which resulted in 20 measurements at each point. These measurements at a seat imitate the dynamic propagation environment and validate the introduction of the statistical method. Seat index was defined as 1 to 144 from the left top to the right bottom seat.

In this experiment, the accuracy of traditional geometry based RSSI, TDOA and the proposed method was tested. In the learning phase, among 20 measurement data at a seat, the cross validation was employed. One recorded signal is picked up and used as the signal from the target node, while the remaining 19 recorded signals were employed for the learning phase to construct the database. As any single among the 20 recorded ones can be selected as the target signal, the proposed method is also tested 20 times per seat. The three methods were evaluated by mean distance error $E[|\hat{u} - u_0|]$.

In Fig. 6, the experiment environment and an example of estimated likelihood map in certain time are given. In Fig. 7, the experiment results are given. The proposed method is obviously accurate in spite of a spatially rough measurement and a forced fitting of the distribution of F into Gaussian distribution. The mean distance error of RSSI and TDOA is up to 7m and it means that they are almost useless considering the size of the class room. In contrast, the proposed method achieves high location estimation accuracy at every seat with a mean distance error much smaller than the seat spacing of 0.78m. This result shows the validity of introducing statistical machine learning method for location fingerprint and rough grid measurement.

## 4. Localization of Wi-Fi devices

In this section, the second application of localization of Wi-Fi device using Wi-Fi training preambles is described [31]. It is especially useful for localization in indoor environments where satellite-based localization systems, e.g. GPS are inadequate due to multipath and signal blockage, but there are large numbers of Wi-Fi devices.

Different from other conventional ones, in the proposed system, RF fingerprints are estimated at the APs based on signals transmitted from Wi-Fi terminals for reducing computation cost on the devices. Furthermore, besides collecting only RSSI information, the proposed system enables the collection of phase difference between different antenna pairs at each AP. Furthermore, CAPWAP (Control and Provisioning of Wireless Access Points), a centralized controlling framework of multiple Wi-Fi APs implemented above the MAC layer of each AP, is utilized to collect RF fingerprints from multiple APs at the APC (AP Controller) [38]. Another important novelty of the proposed system is that autonomous navigation is employed at the surveillance robot to detect its own location to report to the APC in the learning phase, and Pedestrian Dead Reckoning (PDR) is employed at the target Wi-Fi device to enhance the localization accuracy.

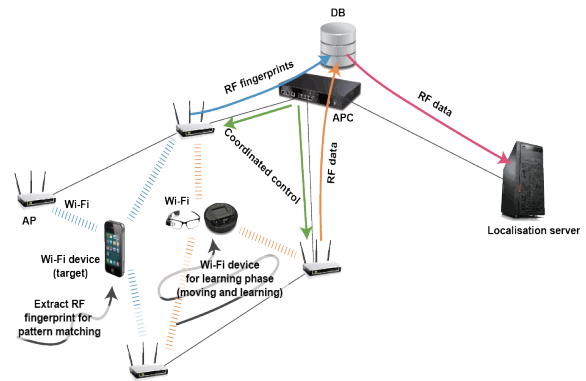

**Fig. 8** System architecture for localization of Wi-Fi devices

## 4.1 System architecture:

The indoor localization system for Wi-Fi devices using multi-sensor (Wi-Fi access points (AP)s) based on fingerprinting technique is shown in Fig. 8. Different from Sect. 3 which requires special hardware, the system in this section is mainly based on commercial products i.e. IEEE802.11a (11a) compliant APs and UEs; except that a central controller called APC is newly developed on top of the deployed multi-sensor APs to aggregate and process fingerprints measured independently at each AP.

The proposed fingerprint-based approach includes two phases: (1) learning phase (2) estimation (localization) phase. In the first phase, a surveillance robot, who can move around in a target indoor environment, is employed to provide RF environment parameters together with its position information to surrounding Wi-Fi APs. Based on the received Wi-Fi signals sent from the robot, the APs extract the propagation parameters e.g. received signal strength and phase difference at different received antennas; and then transfer them to a AP controller. APC collects and stores them in a database called radio map, which is open for application server (AS) to access and achieve sufficient (stochastic) information for localization in the second phase. In the second phase, a target terminal (Wi-Fi device) similarly transmits Wi-Fi signals. APs can listen to these signals and extract propagation parameters and forward them to the APC. Similarly, AS can access the APC and estimate the location of the target terminal by pattern matching between the received parameters of the target device and the constructed radio map in the learning phase.

## 4.2 Localization algorithms:

### A. RF fingerprint

There are many types of RF fingerprints e.g. RSSI, TDOA, AOA, CIR. As many types of fingerprint are collected, the localization accuracy can be improved at the expense of system cost. For example, TDOA requires punctual synchronization between APs. For ease of implementation and reducing cost while scarifying performance compared with Sect. 3 and 6, the proposed system in this section employs only RSSI and AOA-equivalent phase difference information between a pair of received antennas at each AP (RSPD or Received Signal Phase Difference) respectively as follows:

$$\mathbf{F}^{m,(i,j)} = \begin{cases} \frac{1}{N}\sum_{n=1}^{N} y_n^{m,(i)}\left(y_n^{m,(i)}\right)^* \\ arg\left(\frac{1}{N}\sum_{n=1}^{N} y_n^{m,(i)}\left(y_n^{m,(j\neq i)}\right)^*\right) \end{cases} \tag{7}$$

where $y_n^{m,(i)}$ is the received signal sample $n$ at the ($i$) antenna of AP $m$, $N$ is the total number of received samples and $arg()$ denotes the angular operator of a complex number (phase information). In the experiment, the received signal for computing RSSI and RSPD is extracted from the long preambles within the 11a frame. In the above formula, if a same antenna index is employed, $\mathbf{F}^{m,(i,i)}$ denotes the average received RSSI, otherwise $\mathbf{F}^{m,(i,j)}$ denotes the phase difference of a pair of two antennas $(i, j)$ at AP $m$. From now on, let us stack all element of $\mathbf{F}^{m,(i,j)}$ into one common vector $\boldsymbol{F}$ $\forall(m, i, j)$.

### B. Learning phase

In the learning phase, the APC constructs the radio map as the conditional probability density function (PDF) $p_{(\boldsymbol{F}|\mathbf{u})}(\boldsymbol{F}|\boldsymbol{u})$ where $\boldsymbol{u}$ denotes the location vector reported by the mobile robot when measuring $\boldsymbol{F}$, which includes both RSSI and RSPD information. In this paper, a parametric approach is employed to estimate the pdf for RSSI and RSPD respectively. For RSSI, the authors assume Nakagami-$m$ distribution while von-Mises distribution is assumed for phase information based on direction statistics i.e.

$$p_{\Gamma}(x; \beta, \theta) = \frac{x^{\beta-1}}{\Gamma(\beta)\theta^\beta} e^{-\frac{x}{\theta}}$$

$$p_{\text{VM}}(x; \mu, \kappa) = \frac{e^{\kappa\cos(x-\mu)}}{2\pi I_0(\kappa)} \tag{8}$$

where $\Gamma(\cdot)$ is the Gamma function, $\beta$ and $\theta$ are the shape and scale parameters respectively. Also, $I_0(\cdot)$ is the zeroth order

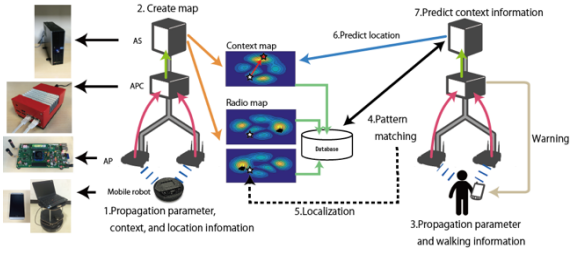

**Fig. 9** Experimental system for indoor localization

Bessel function of the first kind and $\mu$ and $\kappa$ are the average angle and concentration parameters respectively. Therefore, the empirical conditional pdf can be given by the following formula assuming the independence between RSSI and RSPD, i.e.,

$$p_{(\mathbf{F}|\mathbf{u})}(\mathbf{F}|\mathbf{u}) =$$

$$p_{\Gamma}\left(|\mathbf{F}|; \hat{\beta}(\mathbf{u}), \hat{\theta}(\mathbf{u})\right) p_{\mathrm{VM}}\left(arg(\mathbf{F}); \hat{\mu}(\mathbf{u}), \hat{\kappa}(\mathbf{u})\right)$$

where $\widehat{(\cdot)}$ denotes the parametric estimation values of the distributions' characterized parameters based on the measured fingerprints $\mathbf{F}$ at location $\mathbf{u}$.

### C. Estimation (localization) phase:

Similarly to the learning phase, the fingerprint vector $\mathbf{F}^{\mathrm{target}}$ of the target Wi-Fi device for localization can be achieved at the APC. The location of the target device can be estimated as the location maximizing the posterior likelihood function as follows:

$$\hat{\mathbf{u}} = \arg\max_{\mathbf{u}} p_{(\mathbf{F}|\mathbf{u})}(\mathbf{F}^{\mathrm{target}}|\mathbf{u}) \qquad (9)$$

To enhance localization accuracy taking into account the target UE device's movement tracked through its own sensors, i.e. a PDR approach, Eq. (9) can be extended by introducing the pdf of location $p_{(\mathbf{u})}(\mathbf{u})$ given by PDR [39] and approximated by particle filter $\mathbf{u}^{(p=1\,to\,P)}$ where $p$ denotes the particle index, and $P$ denotes the total number of particles.

$$p_{(\mathbf{F}|\mathbf{u})}(\mathbf{F}^{\mathrm{target}}|\mathbf{u}) =$$

$$\sum_{p=1}^{P} p_{(\mathbf{F}|\mathbf{u})}(\mathbf{F}^{\mathrm{target}}|\mathbf{u}^{(p)}) p_{(\mathbf{u})}(\mathbf{u}^{(p)}) \qquad (10)$$

where the likelihood of particle filters are regressed from 4 nearest surrounding training grids $\mathbb{Z}^{(p)}$ i.e. $p_{(\mathbf{F}|\mathbf{u})}(\mathbf{F}^{\mathrm{target}}|\mathbf{u}^{(p)}) = \sum_{i \in \mathbb{C}^{(p)}} \mathbf{w}^{(i,p)} p_{(\mathbf{F}|\mathbf{u})}(\mathbf{F}^{\mathrm{target}}|\mathbf{u}_i)$. Here $\mathbf{w}^{(i,p)}$ denotes the averaging weight determined by specific regression algorithm. In our experiment, this weight is inversely proportional to the distance between the particle and the grid. The particle filters are resampling at each round by removing ones with low likelihood and enhance the samples with high likelihood $p_{(\mathbf{F}|\mathbf{u})}(\mathbf{F}^{\mathrm{target}}|\mathbf{u}^{(p)})$. From (10), it can be seen that the complexity depends on $P$, the number of antenna pairs on each sensor, and the number of sensors $M$. Therefore, we can calculate the complexity as $\mathrm{O}(N_{\mathrm{ant}}^2 MP)$.

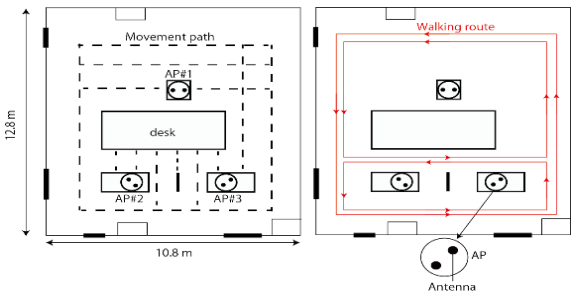

**Fig. 10** Measurement environment and walking routes

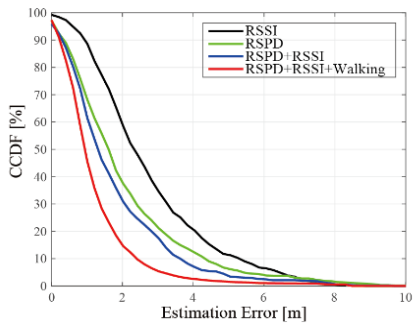

**Fig. 11** Experiment results

### 4.3 Validation experiment:

The localization performance was evaluated by an indoor experiment. The overall experimental system is shown in Fig. 9. In this system [31], three APs each equipped with two antennas were deployed to measure RF fingerprints. The AP was developed based on WARP (Wireless Open-Access Research Platform) v3 Kit wireless board, a product of Mango Communications Inc., in which FPGA firmware was modified to enable the collection of phase difference of elements of the array antenna. The APs support IEEE 802.11a standard with the central frequency of 5GHz. For the APC, Server Taro, a product of PiNON Inc., was employed. For the surveillance robot, Kobuki Turtlebot2, a product of Yujin Robotics Inc., was employed since its movement can be remotely controlled owing to the autonomous navigation function embedded in the robot. Localization algorithm can be executed at the AS through MATLAB programs. Xperia Z1 SOL23 was used as the target of localization, knowing that arbitrary Wi-Fi devices supporting IEEE802.11a can be used.

The experiment was conducted at room I1-754 of Ishikawa-dai campus of Tokyo Institute of Technology as shown in the left-hand side photo of Fig. 10. In the learning phase, the surveillance robot was remotely controlled to move on the routes depicted as dashed lines in this figure to collect required RF fingerprints to construct the radio map at APC. In the estimation (localization) phase, the target Wi-Fi device held by a human was moved on the red routes depicted in the right-hand side photo of Fig. 10. In the experimental system, the information about the terminal's location estimated by itself (PDR) is encapsulated into the DATA field of the packets sent to the APs, at which RF fingerprints of the target device together with its estimated position were collected to perform pattern matching. Detailed information about this Bayesian approach is presented in [40].

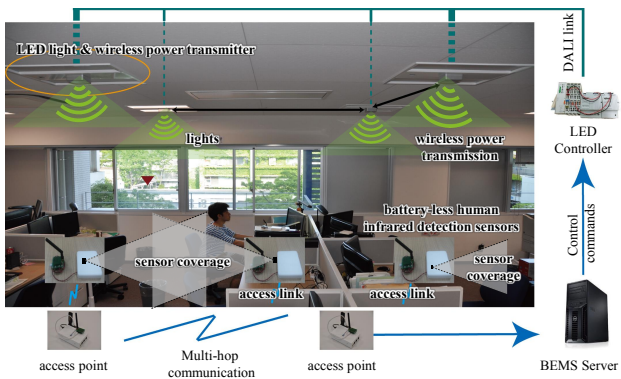

**Fig. 12** Illustration of localization and lighting control system using battery-less sensor network

The experiment result is plotted in Fig. 11, which shows the Complementary Cumulative Distribution Function (CCDF) of localization errors with different combination of fingerprints. It should be noted that all the results in this figure are fingerprint-based methods, while conventional schemes in Fig. 7 of the previous section are geometry based methods. To know the exact location of the terminal, a high resolution camera was used as reference. From this figure, the location accuracy can be ranked as follows: RSSI < RSPD < RSSI+RSPD < RSSI+RSPD+PDR (walking information). Also, the result verified that the best strategy can achieve a mean estimation error 1.36 m. Such high accuracy might be suitable for context-based mmWave beamforming system with outband C-plane [41-43].

## 5. Localization of moving peoples

In this section, a localized light control system in BEMS using location information of peoples estimated by distributed wireless infrared (IR) human detection sensors is introduced [32]. IR sensors have been widely used for localization applications in BEMS due to its high energy efficiency and cost performance, but the poor sensing capability makes it difficult to support accurate localization information for effective control applications. To improve the detection accuracy, the authors locate sensors close to each user by using a battery-less wireless sensor network. A multi-sensors-based localization algorithm is also employed to capture and track user's location more accurately. In the learning phase, detection probability of sensors are derived from actual test, and in the estimation phase, likelihood of each location is calculated iteratively to track the location of users by using a priori knowledge and mobility model. The localization system has been applied in a localized lighting control system, which focuses on reducing lighting energy consumption while satisfying users' illuminance requirement. Two experiments were conducted to verify the performance of localization and lighting control.

### 5.1 System architecture of prototype hardware

The wireless human detection sensor network is implemented in an office [44] for localization, and the location information can be used in lighting control. The overall image and structure of the system are given in Fig. 12.

### A. Human detection sensor network

To address IR sensor's range-only sensibility and low detection accuracy, instead of placing the sensors on the ceiling walls, sensors are flexibly located in working areas close to users by a battery-less wireless sensor network in the target office, as shown in Fig. 12. All employed sensors were battery-less and activated by multiple wireless energy transmitters which are embedded in the ceiling LED lights, as shown in Fig. 13 and Fig. 14. Through a multi-hop communication network, the sensing data are wirelessly sent to the BEMS server, which estimate and track user's location based on the data from distributed sensors.

### B. LED lights and Controllers

The BEMS server estimates user's location and create the LED control commands. All LED lights are connected to a controller, which receives control commands from the BEMS server. A localized illuminance control scheme was considered, because only the illuminance around the users need to be higher than a satisfaction level in an office environment. For that purpose, the optimization problem of illuminance control can be derived:

$$\min_{sw_1, sw_1, \dots} P_{\text{ALL}}$$
$$s.t.$$
$$\begin{cases} I_{\mathbf{u}} \geq I_{\mathbf{u}}^{\text{S}} - I_{\mathbf{u}}^{\text{EN}}, & \forall \mathbf{u} \in \mathbf{U}_t \\ 0 \leq sw_l \leq 1, & \forall l \end{cases} \quad (11)$$

where $sw_l$ is the switch states, $P_{\text{ALL}}$ is the total power consumption of lights, $I_{\mathbf{u}}^{\text{S}}$ denotes the preset illuminance satisfaction level, $I_{\mathbf{u}}^{\text{EN}}$ is the environment illuminance level, and $\mathbf{U}_t$ is a set of locations where luminance requirement should be satisfied. $\mathbf{U}_t$ is given by the localization algorithm in the next section.

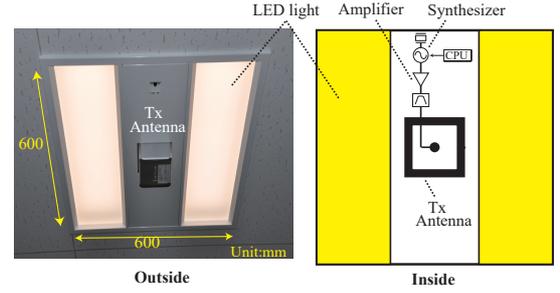

**Fig. 13** LED light and energy transmitter

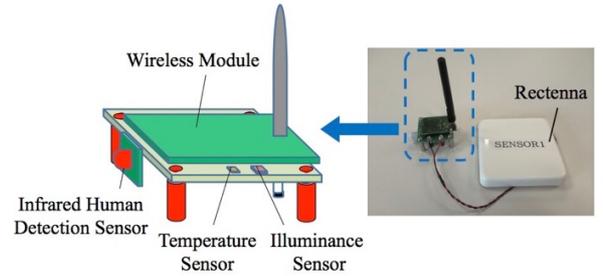

**Fig. 14** Battery-less sensor

### 5.2 User Localization

In this section, a fingerprint based localization algorithm using 1-bit human detection sensors is described. To combat the extremely low resolution of human detection sensors, an iterative user tracking algorithm such as Kalman filter or RNN are introduced.

### A. Learning Phase

In this algorithm, fingerprint vector $\mathbf{F}$ is defined as the output of $M$ human detection sensors $\mathbf{F} = [b_1, b_2, \dots b_M]$. The output of $m$-th sensor $b_{\text{m}}$ is 1 or 0 depending on the relative location between the user and sensor and also the user mobility. Since human detection sensor is detecting differential information of IR signal emitted by people, detection errors happen when the people is completely static. By taking into account the above phenomenon, the conditional probability of m-th fingerprint $p(F_m|\mathbf{u})$ can be described as follows:

$$p(F_m|\mathbf{u}) = \begin{cases} P_{\text{D}}(\mathbf{u}) & \text{for } b_{\text{m}} = 1 \\ 1 - P_{\text{D}}(\mathbf{u}) & \text{for } b_{\text{m}} = 0 \end{cases} \quad (12)$$

where $P_{\text{D}}(\mathbf{u})$ is the detection probability of the sensor depending on the location of user $\mathbf{u}$ by averaging the behavior of users. Assuming the independence of sensors, the conditional probability of fingerprint vector is derived as:

$$p(\mathbf{F}|\mathbf{u}) = \prod_m p(F_m|\mathbf{u}) \quad (13)$$

### B. Estimation phase

In the estimation phase, user tracking is introduced by describing the likelihood function with the learned conditional PDF $p(\mathbf{F}|\mathbf{u})$ and the prior probability of user location $p(\mathbf{u})$:

$$L(\mathbf{u_t}) = p(\mathbf{F}_t|\mathbf{u}_t)p(\mathbf{u}_t)$$
$$= p(\mathbf{F}_t|\mathbf{u}_t)\sum_{\mathbf{u}_{t-1}} p(\mathbf{u}_t|\mathbf{u}_{t-1})p(\mathbf{u}_{t-1}) \quad (14)$$
$$\approx p(\mathbf{F}_t|\mathbf{u}_t)\sum_{\mathbf{u}_{t-1}} p(\mathbf{u}_t|\mathbf{u}_{t-1})\, L(\mathbf{u}_{t-1})$$

where $\boldsymbol{u}_t$ and $\mathbf{F}_t$ are the location of user and fingerprint vector at time $t$ respectively, and $p(\mathbf{u}_t|\mathbf{u}_{t-1})$ is state (location) transition probability of the user. By approximating $p(\mathbf{u}_{t-1})$ with $L(\mathbf{u}_{t-1})$, the estimation and tracking can be done iteratively. The state transition of user, namely mobility model during the period of sensing, is statistically modeled by a two-mode approach to switch between *Static Mode* and *Moving Mode*. In the *Moving Mode*, the motion is assumed to be random walk with independent Gaussian accelerations in both $x$ and $y$ axes.

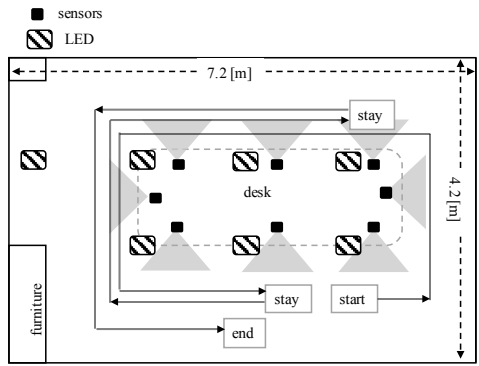

**Fig. 15** Experiments environment

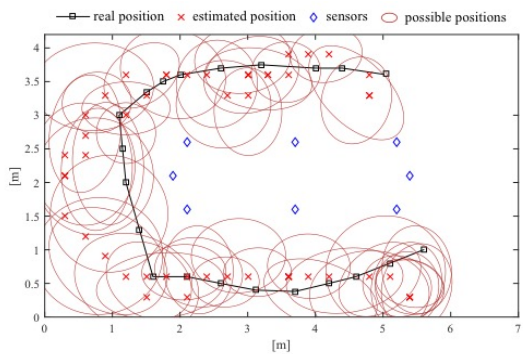

**Fig. 16** MLE localization experiments results

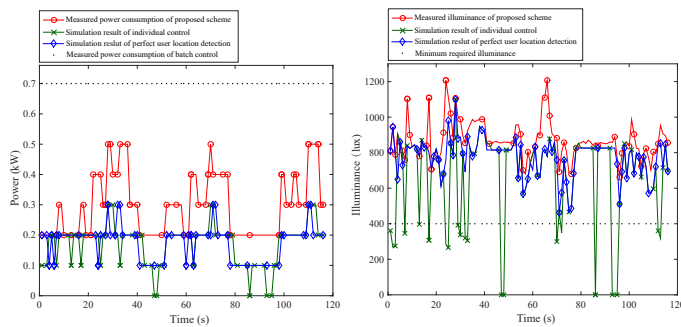

**Fig. 17** Experimental result of power consumption and illuminance

From the above, the location of user can be estimated by the MLE algorithm as in previous sections:

$$\hat{\mathbf{u}}_t = \underset{\mathbf{u}_t}{\mathrm{argmax}}\, L(\mathbf{u}_t) \qquad (15)$$

However, in the application of localized lighting control, estimation error of user location will degrade user satisfaction in terms of luminance. Therefore, not only the location maximizing likelihood, but also the locations with likelihood larger than a threshold are triggered to satisfy user's luminance requirements:

$$\boldsymbol{U}_t = \{\boldsymbol{u}_t | L(\boldsymbol{u}_t) > \eta\} \qquad (16)$$

In the experiment $\eta$ is decided experientially. The set $\boldsymbol{U}_t$ contains all triggered locations.

As shown in (15), localization involves the maximization of the likelihood using a brute-force search. The calculation of the likelihood is based on all historical sensing data, so in Eq. (14) all likelihood in previous time is traversed. Thus, as shown in Table 1, the complexity can be calculated as $\mathrm{O}\left((M + N_{\mathrm{gr}})N_{\mathrm{gr}}\right)$. The calculation can be further reduced if positions with likelihood smaller than a threshold are omitted.

### C. Extension

Currently, the mobility model $p(\mathbf{u}_t|\mathbf{u}_{t-1})$ is statistically given with assumed parameters such as accelerations. However, sometimes the assumed models are not accurate enough and could mismatch with the real world. The machine learning based methods, such as RNN which is good at modeling dynamic temporal process, can be introduced to improve system performance by learning statistical properties of sensing and mobility at the same time from real data, instead of mathematical assumptions.

### 5.3 Experiment

An experiment was conducted to verify the localization performance by the human detection sensor network in an office whose layout is given in Fig. 15. Eight human detection sensors were placed on the desks. In the experiment, the test user walked around in the space, from top right corner to down right corner. The real positions and experimental results, including the positions with maximum likelihood and the positions with likelihood larger than a threshold, are shown in Fig. 16. In the experiment, the localization error was 91cm. Obviously, the performance can be improved by increasing $M$ the number of sensors.

Another localized lighting control experiment was also conducted. The test user walks in the office. The route is shown in Fig. 15 by black lines. When walking, the user measures the illuminance at working surface height, and at the same time, the lighting power consumption is recorded by a power logger. The experiment results are given in Fig. 17. It shows that this localized light control system can reduce the energy consumption significantly, 57%, compared to the batch control scheme, and satisfies users' illuminance requirement with 100% probability.

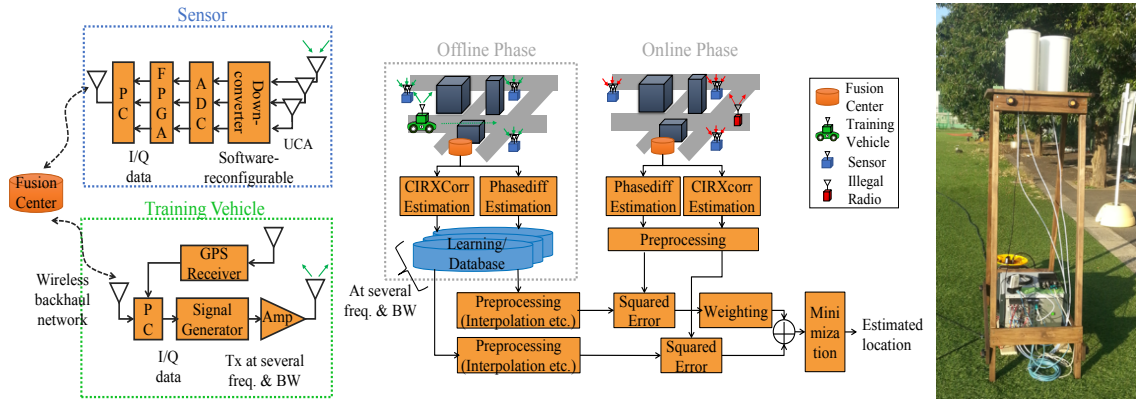

**Fig. 18** System architecture and photograph of developed sensor

## 6. Localization of Radio Terminals Emitting Illegal Radio Signals

In this section, a system to localize unknown illegal radios is introduced [33]. It is crucial to perform localization of illegal radios accurately because they may cause harmful interference to nearby systems which may disrupt important public services. The Ministry of Internal Affairs and Communications (MIC) Japan currently employs the DEURAS-D system [46] to localize illegal radios, which utilizes geometry-based triangulation. However, since triangulation requires Line-of-Sight (LOS), it may result in poor accuracy in urban environments. It is expected that fingerprint-based algorithms would be advantageous for this application because LOS is not required.

However, localization of illegal radios poses several new challenges compared to the applications introduced in the previous sections. Firstly, the localization system does not have any knowledge of the transmitted signal, making it difficult to estimate the channel response as location fingerprints. Secondly, during the offline phase, the localization system also does not have any knowledge regarding the bandwidth, center frequency or the transmit power of the illegal radio, and this makes it difficult to collect fingerprints using the center frequency and bandwidth of the illegal radio in advance. Therefore, it is necessary to learn the fingerprint model from the training data and perform regression on the fingerprints in the bandwidth, frequency and spatial domains.

Considering the application, in this section we will try to address the three questions posed at the end of Sect. 1. Firstly, since we are dealing with illegal radios, we require a fingerprint that can be estimated without knowledge of the transmit signal. Secondly, to enable localization of illegal radios which may appear at any frequency band with an arbitrary bandwidth, we require a software-reconfigurable hardware. Thirdly, to improve the pattern matching results, we require regression and interpolation of the training fingerprints in multiple domains to match the parameters of the illegal radio. In this section, we explain techniques which can overcome the problems above, which were previously introduced [47][48], including a hybrid algorithm to combine the strengths of multiple fingerprints [35]. Simulation results in a dense urban environment are presented, followed by preliminary measurements in an open field conducted using the developed hardware prototype.

### 6.1 System architecture of prototype hardware

The proposed system architecture is shown in Fig. 18. To suit its application for illegal radios, it has several additional components in addition to those shown in Fig. 2. Particularly, the preprocessing block which includes fingerprint interpolation over multiple domains is a crucial new component. Also, unlike the hardware introduced in the previous sections, the RF frontend of the sensors here is software-reconfigurable to cover a wide range of frequency bands from several tens of MHz to several GHz. Furthermore, to learn the fingerprint model for regression in the frequency domain, training signals are transmitted at several frequencies with possibly different bandwidths. A uniform circular array (UCA) is utilized to capture spatial information. GPS receivers are also installed on each Rx sensor for synchronization. The fusion center sends measurement commands to sensors using a wireless backhaul network.

### 6.2 Regression on fingerprints

In order to deal with the problem of unknown transmit signals, the cross-correlation of the received signal at several sensors was utilized as fingerprints [47]. Let the received signal at the $m$-th sensor be denoted as $y_m(t) = h_m(t) * x(t) * g(t) + n(t)$, where $h_m(t)$, $x(t)$ and $g(t)$ denote the CIR, transmit bit sequence and pulse-shaping filter of the transmitter respectively, and $*$ denotes the convolution operation. Then, the cross-correlation between received signals at a single antenna on the $m$-th and $m'$ -th sensors can be expressed as:

$$\mathbb{C}\Big(y_m(t), y_{m'}(t)\Big)$$
$$= \mathbb{C}\Big(h_m(t), h_{m'}(t)\Big) * \mathbb{C}\big(g(t), g(t)\big) + \alpha\delta(t) \quad (17)$$

Here, it is assumed that the transmit bit sequence is random, thus its autocorrelation approximates a delta function. From Eq. (17), it is understood that this fingerprint contains the cross-correlation of CIRs, and unlike the fingerprint in Sec. 3, it can be obtained without CIR estimation or knowledge of the transmitted signal. In this section it will be denoted as Xcorr fingerprints. The cross-correlation is calculated for several delays and is stored as a vector.

Furthermore, in order to utilize the spatial information of the channel, the phase-difference between antenna elements is also utilized as fingerprints [48]. The phase-difference between the $j$-th and $j'$ -th antennas on the $m$-th sensor can be expressed as $\mathbb{C}\Big(y_{j,m}(t), y_{j',m}(t)\Big)$, which is similar to the RSPD fingerprint in Sec. 4. It can be shown that it contains information about the AOA of the dominant multipath. In this section it will be denoted as Phasediff fingerprints, and it is defined separately from Xcorr fingerprints to differentiate the type of information contained in them, and also because their interpolation methods are different.

Due to numerous constraints in the offline phase, it is only possible to collect training fingerprints discretely over space, center frequency and bandwidth. Therefore, regression or interpolation of the training fingerprints should be performed in the spatial, frequency and bandwidth domains before pattern matching is conducted. This is illustrated in Fig. 18 as the additional preprocessing block required before pattern matching.

In the online phase, when this illegal radio appears, spectral estimation techniques are utilized to estimate its center frequency and bandwidth. For Xcorr fingerprints, interpolation in the bandwidth, frequency and spatial domains are required [47]. Bandwidth domain interpolation is performed by filtering the training fingerprints with an appropriate low-pass filter, and this step is crucial for matching of the delay bins. Then, to match the center frequencies, frequency domain interpolation is performed using regression based on a log-linear model.

$$10\log_{10}(F) = \beta_1 \log_{10} f + \beta_2 \quad (18)$$

The left hand side of the equation above denotes a single delay sample of the Xcorr fingerprint in dB scale, and $\beta_1, \beta_2$ denote regression coefficients which are estimated using the training data collected at several frequencies. Next, to increase the density of training fingerprints in the spatial domain, interpolation is performed using Kriging as explained in Sec. 2. Finally, to deal with the unknown transmit power, all fingerprints are normalized by the largest Rx power among all sensors.

For Phasediff fingerprints, interpolation in the frequency and spatial domains are required [48]. Frequency domain interpolation is performed by estimating the dominant multipath's AOA and using it in the array response at the illegal radio's frequency. Linear interpolation is employed for spatial interpolation, and an additional weighting step was introduced to compensate fingerprints where there is no dominant multipath.

A maximum likelihood approach was employed to derive a hybrid algorithm which combines the strengths of both fingerprints [35]. Since the Xcorr and Phasediff fingerprints are statistically independent, localization can be performed by maximizing their joint log-likelihood as the following equation:

$$\hat{\mathbf{u}} = \arg\max_{\mathbf{u}_i}\Big[\log(L^{\text{Xcorr}}(\mathbf{u}_i)) + \log\Big(L^{\text{Phasediff}}(\mathbf{u}_i)\Big)\Big] \quad (19)$$

If we assume that the likelihood function of both fingerprints can be modeled as a zero mean Gaussian distribution with different variance, Eq. (19) can be rewritten in terms of the squared error $\varepsilon_i^2$ between the fingerprints of the illegal radio and the interpolated training fingerprints corresponding to coordinates $\mathbf{u}_i$, and also the measurement error variance $\sigma^2$ for each type of fingerprint.

$$\hat{\mathbf{u}} = \arg\min_{\mathbf{u}_i}\left[\frac{\left(\varepsilon_i^{\text{Xcorr}}\right)^2}{2\left(\sigma^{\text{Xcorr}}\right)^2} + \frac{\left(\varepsilon_i^{\text{Phasediff}}\right)^2}{2\left(\sigma^{\text{Phasediff}}\right)^2}\right]$$
$$= \arg\min_{\mathbf{u}_i}\left[\left(\varepsilon_i^{\text{Xcorr}}\right)^2 + \gamma\left(\varepsilon_i^{\text{Phasediff}}\right)^2\right] \quad (20)$$

$\gamma$ denotes the ratio between error variance of the two fingerprints. In practice, it would be difficult to estimate the error variances, thus in this paper, $\gamma$ was optimized heuristically.

Optimization in Eq. (20) can be performed using the brute-force search. The Xcorr fingerprint is similar to that in Sect. 3, and the Phasediff fingerprint is similar to the RSPD fingerprint in Sect. 4. Thus, the complexity is roughly on the same order

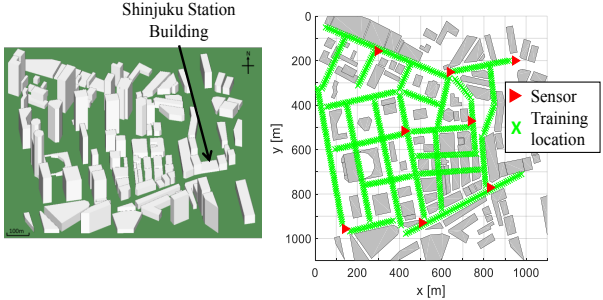

Shinjuku Station
Building

Sensor
Training
location

**Fig. 19** 3D model and map of environment used in simulations.

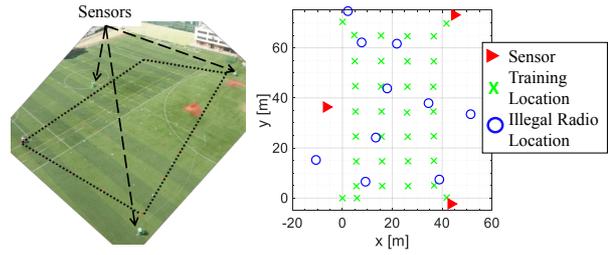

Sensors

Sensor
Training
Location
Illegal Radio
Location

**Fig. 21** Map of environment in experiments

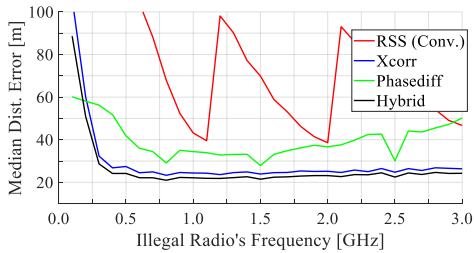

**Fig. 20** Comparison of localization schemes through simulations.

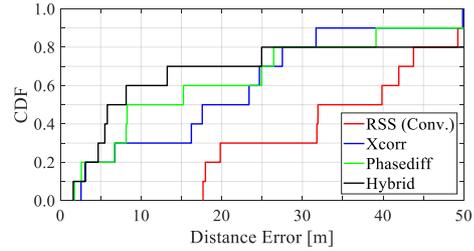

**Fig. 22** Comparison of localization schemes through experiments.

as both algorithms combined. However, in this algorithm, there are additional interpolation steps in the bandwidth, frequency and spatial domains. These steps cannot be performed beforehand in the offline phase because the signal parameters of the illegal radio can only be estimated or known after it has appeared in the area of interest. It is difficult to give a simple estimate of the total complexity, but details regarding each step can be found in [48] [49]. Nevertheless, the complexity can be greatly reduced by employing particle filters instead of the brute-force search which requires interpolation of the whole database. This enables us to sequentially interpolate small subsets of the training database depending on the location of the particles. Details can be found in [50], and are omitted here due to space limitations.

### 6.3 Simulation Results

Ray-tracing simulations were conducted using a model of the 1km² area surrounding Shinjuku station, Tokyo, as shown in Fig. 19. Training fingerprints from 1616 locations with 5m spacing were collected along the main roads at three training frequencies 0.8GHz, 1.5GHz and 2.5GHz with a 10MHz bandwidth and 27dBm Tx power. Localization accuracy was examined with illegal radios placed at a total of 1232 location on a grid with 20m spacing with 5MHz bandwidth and 30dBm Tx power. Detailed simulation parameters can be found in [48].

Figure 20 shows a comparison of the median distance error between various localization algorithms over a wide range of illegal radio frequencies. The hybrid algorithm achieved the best localization accuracy with median distance error of about 22m in the upper frequency range. We can also see that the localization performance of the conventional RSSI algorithm greatly depends on frequency, and is best only near the training frequencies. On the other hand, due to the interpolation in the frequency domain, the proposed techniques achieved roughly the same performance regardless of frequency in the upper frequency range.

### 6.4 Measurement Results

A preliminary experiment was conducted in an open field in the Ōokayama campus of Tokyo Institute of Technology using a developed hardware prototype. Three sensors were placed on the perimeter of the targeted area of 50m x 70m, as shown in Fig. 21. A UCA with 3 omnidirectional antenna elements was utilized at a height of 1.5m. The sensors were able to cover a wide frequency range from 20MHz to 3GHz with a sample rate of 100Msps. Training fingerprints were collected from a total of 33 locations with 10m spacing at frequencies 438.5MHz and 2486.5MHz, with bandwidths of 1MHz and 5MHz, and Tx powers of 0.5W and 0.1W, respectively. In the online phase, the training system was placed at 10 random locations in the targeted area at 1297MHz frequency, 1MHz bandwidth and 0.5W transmit power in order to mimic the operation of unknown illegal radios. More details can be found in [35].

Figure 22 shows a comparison of the CDF of distance error between several localization algorithms. In this experiment,

it is found that there was little advantage in using the Xcorr fingerprints as there were very few multi-paths which could be exploited for localization in the LOS environment. However, by utilizing the hybrid algorithm, an improvement in median distance error of about 30m can be achieved, compared the conventional technique.

## 7 Conclusion

The main contribution of this paper is twofold: 1. The authors discussed the logic and feasibility of fingerprint selection for applications with different typical requirements and limitations, which are summarized by the flowchart in Fig.1. Readers can follow a similar logic process to choose the most suitable and feasible fingerprints for their own applications. 2. The algorithm selection based on the required localization accuracy for the specific application is also a key point, since different algorithms could result in totally different accuracy of performance and hardware complexity. Thus, the accuracies and hardware complexities of representative fingerprint-based localization algorithms are also discussed in details based on four practical applications. These are very important references and baselines for system designers to choose and design the appropriate localization algorithms and estimate the corresponding hardware cost.

Rather than purely from the viewpoint of theoretical algorithms, in this paper, the authors discussed both the algorithms design and hardware details of four different approaches of fingerprint-based radio emitter localization systems as examples of practical applications. Four different typical applications selected in this paper were: localization of LTE smart phones used for anti-cheating in exams, indoor localization of Wi-Fi terminals, localized light control in BEMS using location information of occupants, and illegal radio localization in outdoor environments.

In the future, the following directions may lead the research to further increase the localization accuracies and extend the applications. For instance, combining fingerprint-based localization with other external sensors, such as Pedestrian Dead Reckoning (PDR) sensors and vision sensors, can complement each other's disadvantages and will increase the accuracies, and machine learning is supposed not only to reduce the computational complexities, but also to mine new hidden features and relations between fingerprints and locations to achieve better performance.

It is expected that readers, who plan to build fingerprint-based localization systems, can use this paper as a guideline for both algorithm design and hardware implementation in varieties of environments (e.g., office, classroom and urban area) and for varieties of applications (e.g., radio terminal detection and location-based BEMS).

## Acknowledgments

The authors would like to thank Shintaro Arata, Daisuke Hayashi, Tsutomu Mitsui, Toshihiro Yamaguchi from Koden Electronics for their continuous financial support, and also in developing the hardware and conducting measurements.

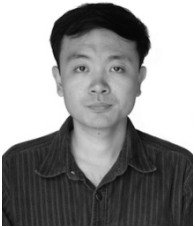

**Tao Yu**     received the B.E. degree in communication engineering from Taiyuan Institute of Technology, China, in 2008, the M.E. degree in signal and information processing from Communication University of China, in 2010, and Dr.Eng. degree in electrical and electronic engineering from Tokyo Institute of Technology in 2017, where he is working as a postdoctoral researcher. His research interests are sensor networks, control and building energy management. He is a member of IEICE.

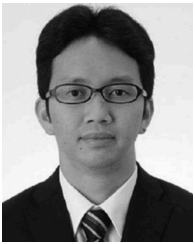

**Azril Haniz**     received the B.E. degree in electrical and electronic engineering in 2010 and the M.Eng and Dr.Eng. degrees from the Department of International Development Engineering, Tokyo Institute of Technology, Tokyo, Japan, in 2012 and 2016, respectively, where he is currently working as a specially appointed Lecturer. His research interests include localization, cognitive radio, sensor networks, and signal processing. He received the best student paper award in the Singapore–Japan International Workshop on Smart Wireless Communications (SmartCom) in 2014 and received the 2016 Tejima Seiichi Doctoral Dissertation Award. He is currently a member of IEICE.

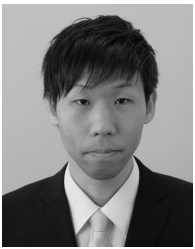

**Kentaro Sano**     received the B.E. and M.E. degree in electrical and electronic engineering from Tokyo Institute of Technology, Japan, in 2012 and 2014 respectively. In 2014, he joined Sony Corporation, Tokyo, Japan and shift to Qoncept, Inc., Tokyo, Japan in 2017. He has been involved research and product development work of image processing and recognizing.

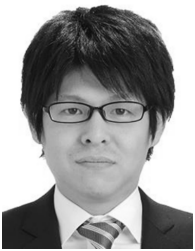

**Ryosuke Iwata**     received the B.E. and M.E. degrees in electrical and electronic engineering from Tokyo Institute of Technology in 2013 and 2015 respectively. He has been engaged in the research and development of cognitive MIMO radio system. His research interests are MIMO cognitive radio and localization using electromagnetic wave.

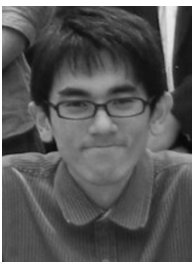

**Ryouta Kosaka**     received the B.E. and M.E. degrees in electrical and electronic engineering from Tokyo Institute of Technology in 2014 and 2017 respectively. He has been engaged in the research and development of localization using electromagnetic wave for context information prediction in indoor environment. His research interests are MU-MIMO and localization using electromagnetic wave. He had received the Young Researcher of the Year Award in Smart Radio from the IEICE SR technical committee and the Best Paper Award in the 2nd CQ workshop from the IEICE CQ technical committee both in 2016.

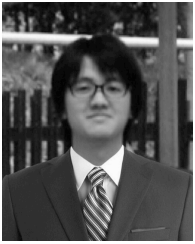

**Yusuke Kuki**    received the B.E. degree in electronic and information engineering from Osaka University, Osaka, Japan in 2013, and the M.E. degree in electrical, electronic and in- formation engineering from Osaka University, Osaka, Japan in 2016.

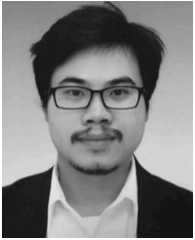

**Gia Khanh Tran**    received the B.E., M.E., and D.E. degrees in electrical and electronic engineering from Tokyo Institute of Technology, Tokyo, Japan, in 2006, 2008, and 2010, respectively, where he is currently an Assistant Professor. His research interests include signal processing, multiple-input multiple-output mesh networks, coordinated heterogeneous cellular networks, millimeter-wave communication, and localization. He received the IEEE VTS Japan Young Researchers Encouragement Award from the IEEE VTS Japan Chapter in 2006 and the IEICE Service Recognition Awards in 2013 and 2015. He also received the Best Paper Award in Software Radio from the IEICE SR technical committee in 2009 and 2013, the Best Paper Award at SmartCom2015, and the Best Paper Awards from both IEICE and IEICE ComSoc in 2014. He served as a Technical Program Committee co-chair in a series of IEEE WDN workshops, including WDN-5G in ICC2017. He is currently the Assistant of the technical committee on Smart Radio of the IEICE ComSoc. He is a member of IEEE.

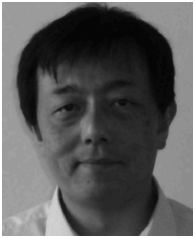

**Jun-ichi Takada**    received the B.E. and D.E. degrees from Tokyo Institute of Technology (Tokyo Tech), Tokyo, Japan, in 1987 and 1992, respectively. He was a Research Associate at Chiba University during 1992–1994 and an Associate Professor at Tokyo Tech during 1994–2006. He has been a Professor with Tokyo Tech since 2006. During 2003–2007, he was also a Researcher at the National Institute of Information and Communications Technology. He served as a Secretary and the chair of IEICE Technical Committee on Software Radio during 2001–2007 and 2007–2009, respectively. His current interests include the radiowave propagation and channel modeling for various wireless systems and regulatory issues of spectrum sharing. He received the Achievement Award of IEICE in 2009. He is a member of Japan Society for International Development.

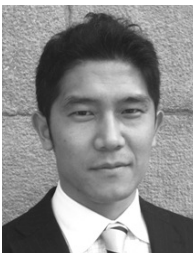

**Kei Sakaguchi**    received the M.E. degree in Information Processing from Tokyo Institute of Technology in 1998, and the Ph.D degree in Electrical & Electronics Engineering from Tokyo Institute Technology in 2006. Currently, he is a Professor at Tokyo Institute of Technology in Japan and at the same time he is working at Fraunhofer HHI in Germany as a Senior Scientist. He received the Outstanding Paper Awards from SDR Forum and IEICE in 2004 and 2005 respectively, and three Best Paper Awards from IEICE communication society in 2012, 2013, and 2015. He also received the Tutorial Paper Award from IEICE communication society in 2006. He served as a General co-chair in the IEEE WDN-5G in 2017 and an Industrial Panel co-chair in the IEEE Globecom in 2017. His current research interests are in 5G cellular networks, millimeter-wave communications, and wireless energy transmission. He is a member of    IEEE.